\documentclass[prl,twocolumn,floatfix, showpacs]{revtex4}
\usepackage{graphicx}
\usepackage{amsmath}
\usepackage{bm}

\begin{document}

\title{Pseudospin valve in bilayer graphene: towards graphene-based pseudospintronics}
\author{P. San-Jose}
\author{E. Prada}
\author{E. McCann}
\author{H. Schomerus}
\affiliation{Department of Physics, Lancaster University,
Lancaster, LA1 4YB, United Kingdom}

\date{\today}
\begin{abstract}
We propose a non-magnetic, pseudospin-based version of a spin
valve, in which the pseudospin polarization in neighboring regions
of a graphene bilayer is controlled by external gates. Numerical
calculations demonstrate a large on-off ratio of such a device.
This finding holds promise for the realization of
pseudospintronics: a form of electronics based upon the
manipulation of pseudospin analogous to the control of physical
spin in spintronics applications.
\end{abstract}
\pacs{75.70.Ak, 75.47.Pq, 85.75.-d}
\maketitle

\begin{figure}[top]
\includegraphics[width=0.9\linewidth]{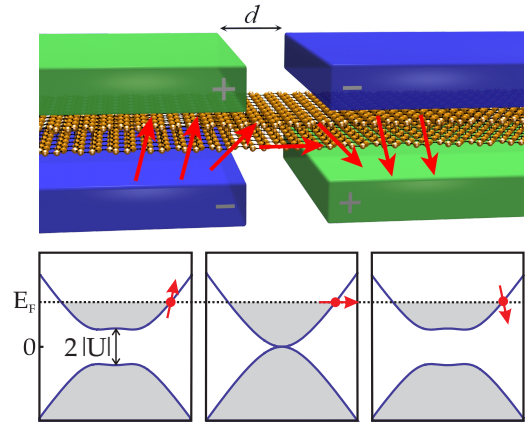}
\caption{\label{fig:1} Pseudospin-valve effect in bilayer
graphene. Schematic diagram of a pseudospin valve in bilayer
graphene in its antiparallel (AP) configuration.}
\end{figure}

Soon after its discovery \cite{novo04}, it was realized that
graphene supports an additional quantum number called pseudospin
\cite{novo05,zhang05,geimrev} that arises because the honeycomb
lattice is composed of two triangular sublattices. Wave-function
amplitudes can be written like the two components of a spin-$1/2$
elementary particle, and electrons in graphene display
characteristics analogous to relativistic fermions
\cite{novo05,zhang05}. In particular, this includes the celebrated
effect of chirality, whose profound consequences include an
unusual sequencing of plateaus in measurements of the quantum Hall
effect \cite{novo05,zhang05,novo06}, suppression of backscattering
\cite{ando98,mceuen99} and Klein tunneling at interfaces
\cite{kats06,chei06}.

So far, it has not been possible to exploit the pseudospin degree
of freedom in graphene in a similar way as physical spin in
spintronics \cite{spintronics} and quantum computing \cite{loss98}
applications. In a monolayer of graphene, chirality means that the
orientation of an electron's pseudospin is inextricably linked to
the direction of its momentum, thus constraining the pseudospin to
lie in the plane of the graphene sheet and preventing its use as
an independently-tunable degree of freedom. In bilayers of
graphene \cite{novo06,mcc06,ohta06}, the pseudospin degree of
freedom is associated with the electronic density on the two
layers. The constraint of chirality entails that electronic
density is equally divided between the two layers so that the
pseudospin again lies in the plane of the layers but now turning
twice as quickly as the direction of momentum \cite{novo06,mcc06}.
Min {\em et al} \cite{min08} were the first to realize that
bilayers still offer a  promising platform for pseudospintronics.
In particular, they predicted that a pseudomagnetic state can form
spontaneously due to strong Coulomb interactions at vanishing
charge-carrier density; this effect is intimately tied to the fact
that the density of states remains finite because  the dispersion
relation is parabolic, in contrast to the situation in a
monolayer.

In this Letter we propose a variant of graphene-based
pseudospintronics which exploits another direct advantage of a
bilayer over a monolayer, namely the facility to induce a
difference between the on-site energies on the two layers via a
perpendicularly applied electric field, which can be realized by
pairs of gate electrodes (see Fig.\ \ref{fig:1}). The resulting
asymmetry of the layers induces an energy gap between the
conduction and valence bands \cite{mcc06,guinea06,mcc06b,min07},
as observed in photoemission \cite{ohta06} and transport
\cite{oostinga,castro} measurements. For states above or below the
gap, interlayer asymmetry has the effect of creating an ``up'' or
``down'' component of pseudospin perpendicular to the electronic
momentum and the plane of the sheet \cite{mcc06,min08}. The
electric field hence acts on the pseudospin in the same way as a
magnetic field acts on the physical spin of electrons in
spintronic applications. In particular, the preferred pseudospin
direction can be switched by inverting the sign of the applied
potential difference. In analogy to the giant magnetoresistance
(GMR) induced by a domain wall boundary in magnetic materials
\cite{gmr,viret}, one would therefore expect that interfaces
between regions of different gate polarity inhibit the flow of
electrons. We will demonstrate that this effect can indeed be
utilized to realize an all-electronic, pseudospin-based analogue
of a spin valve with a large on-off ratio.

\begin{figure}[tb]
\centerline{\includegraphics[width=0.9\linewidth]{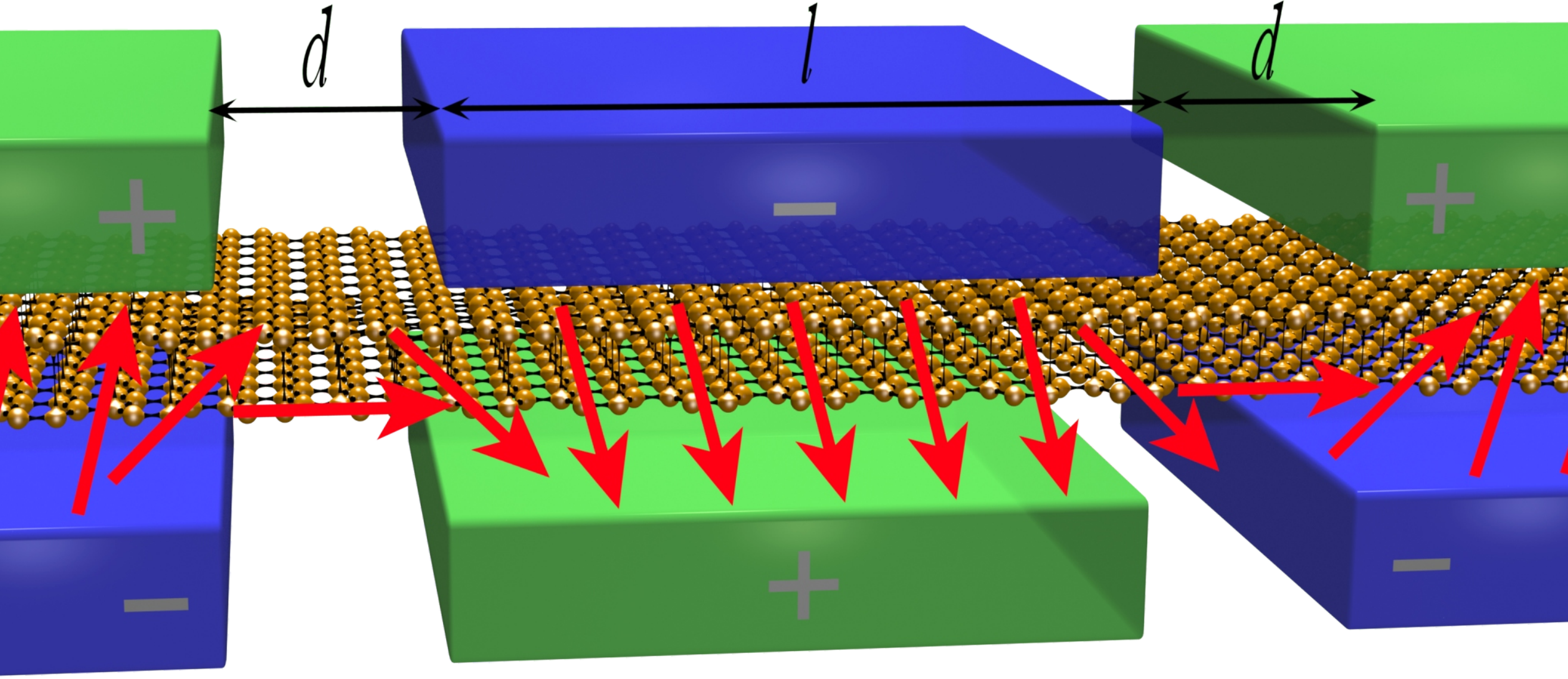}}
\caption{\label{fig:2a} Schematical illustration of a
pseudospin-valve transistor.  This device is operated by switching
the polarity of a central gate of length $l$ (shown is the
antiparallel configuration).}
\end{figure}

{\em Concepts.---}The proposed pseudospin valve can be realized in
a $2D$ sheet of bilayer graphene with sets of gates which produce
a spatial variation of interlayer asymmetry in the direction of
current flow (see Fig.~\ref{fig:1}). Top and bottom gates are used
to independently control the Fermi level and the interlayer
asymmetry, the latter creating an out-of-plane component of
pseudospin. When the polarity of the two pairs of gates is
identical, the device is in its ``parallel'' configuration, and
offers only a small resistance to the flow of electrons with
energies above the gap. The illustration in Fig.\ \ref{fig:1}
shows the device in its ``anti-parallel'' configuration, which is
realized when the polarity of the gates changes sign across the
device. This produces a corresponding rotation of the pseudospin
polarization with a switching of the out-of-plane component.
Similarly to spin scattering at domain walls \cite{viret}, the
pseudospin of an incoming electron will precess about the changing
local polarization as it attempts to follow it. If the change in
the polarization rotation is sharp enough, the re-alignment of the
electron's pseudospin should only be partially successful, leading
to reflection and a drop in the flow of current through the
device.

We characterize the fidelity of the pseudospin valve in terms of
the pseudo-magnetoresistance ($\rm PMR$) ratio
\begin{eqnarray} \label{PMR}
{\rm PMR}=\frac{R_{\rm AP}-R_{\rm P}}{R_{\rm AP}} \, ,
\end{eqnarray}
which is defined by the resistances $R_{\rm P}$ (parallel
configuration) and $R_{\rm AP}$ (antiparallel configuration)
determining the current $I=V/R$ flowing through the device in
response to an applied bias voltage difference $V$. The $\rm PMR$
resistance ratio is the analogue of the conservative definition of
magnetoresistance in spintronic applications
\cite{spintronics,viret}, and takes the value $100\%$ for a
perfect spin valve.

As in conventional spintronic applications, we also consider how
the pseudospin-valve effect can be extended to a broader range of
energies via serial connection of regions of different polarity.
This leads to the design of a pseudospin-valve transistor operated
by switching the polarity of a central gate of length $l$
(shown in Fig.\ \ref{fig:2a}).

\begin{figure}[tb]
\includegraphics[width=0.9\linewidth]{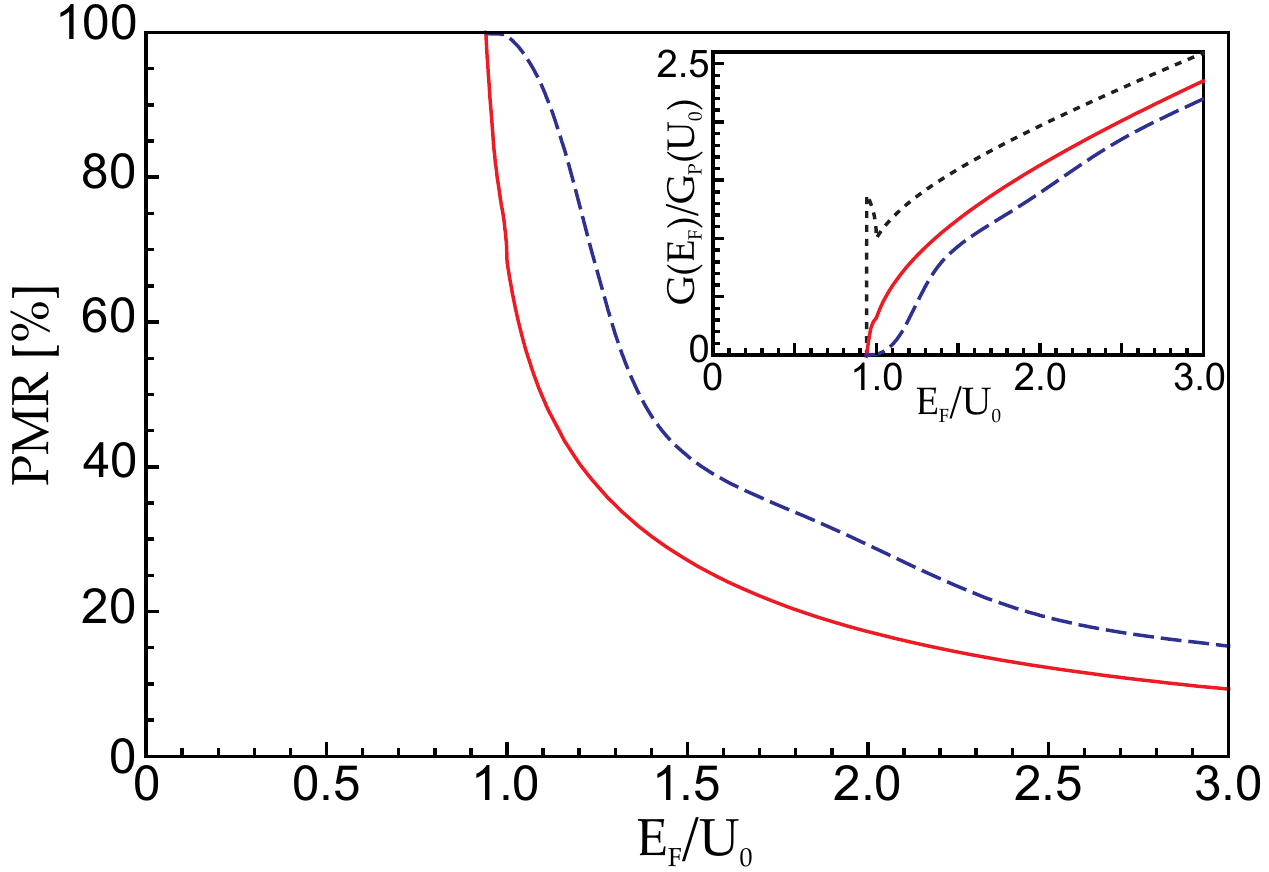}
\caption{\label{fig:PMR} Pseudo-magnetoresistance ratio ${\rm PMR}
= (R_{\rm AP} - R_{{\rm P}})/R_{\rm AP}$ of the pseudospin valve
(where P refers to the parallel configuration) versus the Fermi
level of incoming electrons for $d = 50a$ (solid curve). Here,
$U_0=0.07 eV$ is the magnitude of the gate potential at large
distances. The dashed curve refers to the pseudospin-valve
transistor shown in Fig.~\ref{fig:2a}, with $l=50 a$. Inset:
Conductance versus Fermi energy in the antiparallel (solid and
dashed curves) and parallel (dotted curve) configurations from
which the $\rm PMR$ is derived. The normalization factor is
$G_{\rm P}(E_F=U_0)={4U_0W}/({3\pi a\gamma_0})$.}
\end{figure}

{\em Numerical results.---}We start with numerical results, which
are based on the microscopic tight-binding model of bilayer
graphene. This model offers an accurate description of electronic
transport in terms of a small number of characteristic energies
and length scales. Transport between carbon atoms in a single
layer (arranged on a honeycomb lattice with bond length
$a=1.42\,$\AA) is described by kinetic hopping energy
$\gamma_0\approx 2.9$ eV, which also determines the Fermi velocity
$v_0 =\left( 3/2\right) a \gamma_{0}/\hbar$ of an isolated
monolayer. In a bilayer, the two sheets of carbon are arranged
according to Bernal stacking, whereby half of the atoms are
strongly coupled to an atom in the other layer, with a strength
determined by the interlayer coupling parameter $\gamma_1\approx
0.39$ eV. Additional next-nearest neighbor couplings are
non-essential for the problem at hand, and are therefore neglected
for simplicity.

In the parallel configuration, the spatially constant on-site
potential takes the value $U_{\rm top}=U_0$ in the top layer and
$U_{\rm bottom}=-U_0$ in the bottom layer. In the anti-parallel
configuration of the device we model the on-site potential by
\begin{eqnarray} \label{Uprofile}
U_{\rm top}(x) =-U_{\rm bottom}(x)=U(x)\equiv - U_0 \, {\rm erf} (x / d),
\end{eqnarray}
where ${\rm erf}$ is the error function and $x$ is the coordinate
in the direction of transport. The main design parameters
of the pseudospin valve are the typical length scale $d$ of
variation of the gate potential and the magnitude $U_0$ of the
potential value at large distance, as well as the Fermi energy
$E_F$ which determines the energy at which the electrons are
injected from the electrodes ($E_F=0$ for a charge-neutral gapless
bilayer). In the parallel configuration, the presence of a
homogeneous symmetry-breaking onsite potential opens an energy gap
$2|U_0|/\sqrt{1+(2U_0/\gamma_1)^2}$ around the Fermi energy of the
charge-neutral bilayer (see again Fig.\ \ref{fig:1}). For the
antiparallel configuration with the inhomogeneous potential $U(x)$
of Eq. (\ref{Uprofile}), the solution of the tight-binding model
requires, in general, a numerical approach.

Since $U$ is $y-$independent, the problem of an infinitely wide
ribbon is separable, and electronic modes with fixed transverse
wavenumber $k_y$ decouple. For each transverse mode, the problem
can be reduced to a one-dimensional chain of coupled bilayer unit
cells, where each unit cell is composed of four carbon atoms. The
Green function of each chain can be computed efficiently using the
recursive Green function technique \cite{datta}, which delivers
the transmission amplitude $t(k_y)$ via the Fisher-Lee formula
\cite{fisherlee}. In the linear response regime, the total
phase-coherent conductance of the nanoribbon is then obtained from
the Landauer formula. For a ribbon of finite width $W$, the
following considerations remain valid as long as $W\gg d$ and $k_F
W\gg 1$, so that the contribution of edges can be neglected (see
the discussion at the end of the paper).

The calculated conductance versus Fermi energy is shown in the
inset of Fig. \ref{fig:PMR} for the parallel configuration, as
well as the antiparallel configuration with one or two interfaces.
In these calculations, the interface parameters are $d=50\,a$ and
$d=l=50\,a$, respectively, and the asymptotic gap is $U_0=0.07$
eV, corresponding to parameters which can be realized in present
bilayer experiments \cite{ohta06,castro}. We find that close to
the band edge the conductance in the antiparallel configuration is
strongly reduced below its value in the parallel configuration.
The resulting $\rm PMR$ ratio is plotted in the main panel of Fig.
\ref{fig:PMR}. For energies just above the gap, the $\rm PMR$
peaks at $100\%$. For increasing Fermi energy the resistance ratio
drops, which can be attributed to the decreasing out-of-plane
component of the pseudospin of incoming electrons as they become
less sensitive to the asymmetry of the layers when their kinetic
energy increases. As expected from the GMR analogy, the presence
of a second interface in the  pseudospin-valve transistor (dashed
line) significantly extends the energy range over which the ${\rm
PMR}$ is $\approx 100\%$.

{\em Analytical considerations.---}A qualitative analysis of the
pseudospin-valve effect can be achieved by considering the
low-energy physics of gapped graphene bilayers. The microscopic
tight-binding Hamiltonian delivers a band structure with four
bands. For realistic values of the charge carrier density, the
Fermi surface of bilayer graphene lies in the vicinity of two
valleys, indexed by $\xi=\pm 1$, situated at the K and K' point at
the corners of the hexagonal Brillouin zone. Owing to the
interlayer coupling, two of the bands are split away by an energy
$\approx\pm\gamma_1$. For the interlayer asymmetries and Fermi
energies $|U_0|, |E_F|\ll \gamma_1$ assumed in this work, these
split bands do not contribute to the electronic transport. In the
absence of layer asymmetry, the two remaining bands touch at zero
energy and have an approximately parabolic dispersion relation $E
\approx \pm p^2 /2m$, with effective mass $m = \gamma_{1} /
2v_0^{2}$ and corresponding Fermi velocity $v_F \approx 2 v_0
\sqrt{|E|/|\gamma_1|}$.

To explain the influence of interlayer asymmetry, we employ a
two-component Hamiltonian \cite{mcc06} that approximately
describes the electronic behavior in these two low-energy bands,
\begin{eqnarray*}
H_2 \approx -\frac{1}{2m} \left(
\begin{array}{cc}
0 & \left( \xi{p}_{x}-i{p}_{y}\right) ^{2} \\
\left( \xi{p}_{x}+i{p}_{y}\right) ^{2} & 0
\end{array}
\right) + \left(
\begin{array}{cc}
U & 0 \\
0 & - U
\end{array}
\right) .
\end{eqnarray*}
The effective Hamiltonian $H_2$ operates in a space of
two-component wave functions $\Psi$ describing electronic
amplitudes on the top and bottom layers. The first term in the
Hamiltonian corresponds to a pseudospin-orbit coupling and ensures
chirality of the electronic states in the absence of a
symmetry-breaking on-site potential. The second term in $H_2$
takes into account the influence of external gates that produce
different on-site energies $\pm U$ on the two layers. This term is
analogous to the Zeeman energy of a physical spin in a magnetic
field parallel to the $z$ direction and leads to a gap $2|U|$ in
the electronic spectrum $E_{\pm} \approx \pm \left[ U^2 +
\left(p^2 / 2m \right)^2\right]^{1/2}$. The pseudospin part of the
corresponding wave functions takes the form
\begin{eqnarray*}
\Psi_{\pm} = \frac{1}{\sqrt{2}} \left( \begin{array}{c}
\sqrt{1 + U /E} \,\, e^{-i\xi\phi} \\
\mp \sqrt{1 - U /E} \,\, e^{i\xi\phi}
\end{array}
\right) \, ,
\end{eqnarray*}
where $\phi$ is the angle of the momentum in the plane ${\mathbf
p} = (p \cos \phi , p \sin \phi)$. The pseudospin of such a state
is
\begin{eqnarray*}
\langle{\boldsymbol\sigma}\rangle =  \mp \sqrt{1 - \left( \frac{U}{E}\right)^2}
\left( {\hat \imath} \cos 2\phi  + {\hat \jmath} \, \xi \sin 2\phi
\right) + {\hat k} \frac{U}{E} \, .
\end{eqnarray*}
For energy near the vicinity of the gap $|E| \approx |U|$, the
out-of-plane component takes its maximum value $\langle \sigma_z
\rangle\approx 1$, whereas it is reduced away from the gapped
region.

The pseudospin-valve effect proposed in the present paper
originates in the large resistance at interfaces between regions
of opposite preferred pseudospin direction. This resistance arises
because the pseudospin degree of freedom can adjust itself to such
a spatial variation only over a distance $l_s = h v_0
/\sqrt{|E\gamma_1|}$, as follows from the scaling of the different
terms in the two-component model. This pseudospin precession
length scale is comparable to the Fermi wavelength, which is the
scale on which chirality is established in the symmetric bilayer.

\begin{figure}[tb]
\centerline{\includegraphics[width=0.9\linewidth]{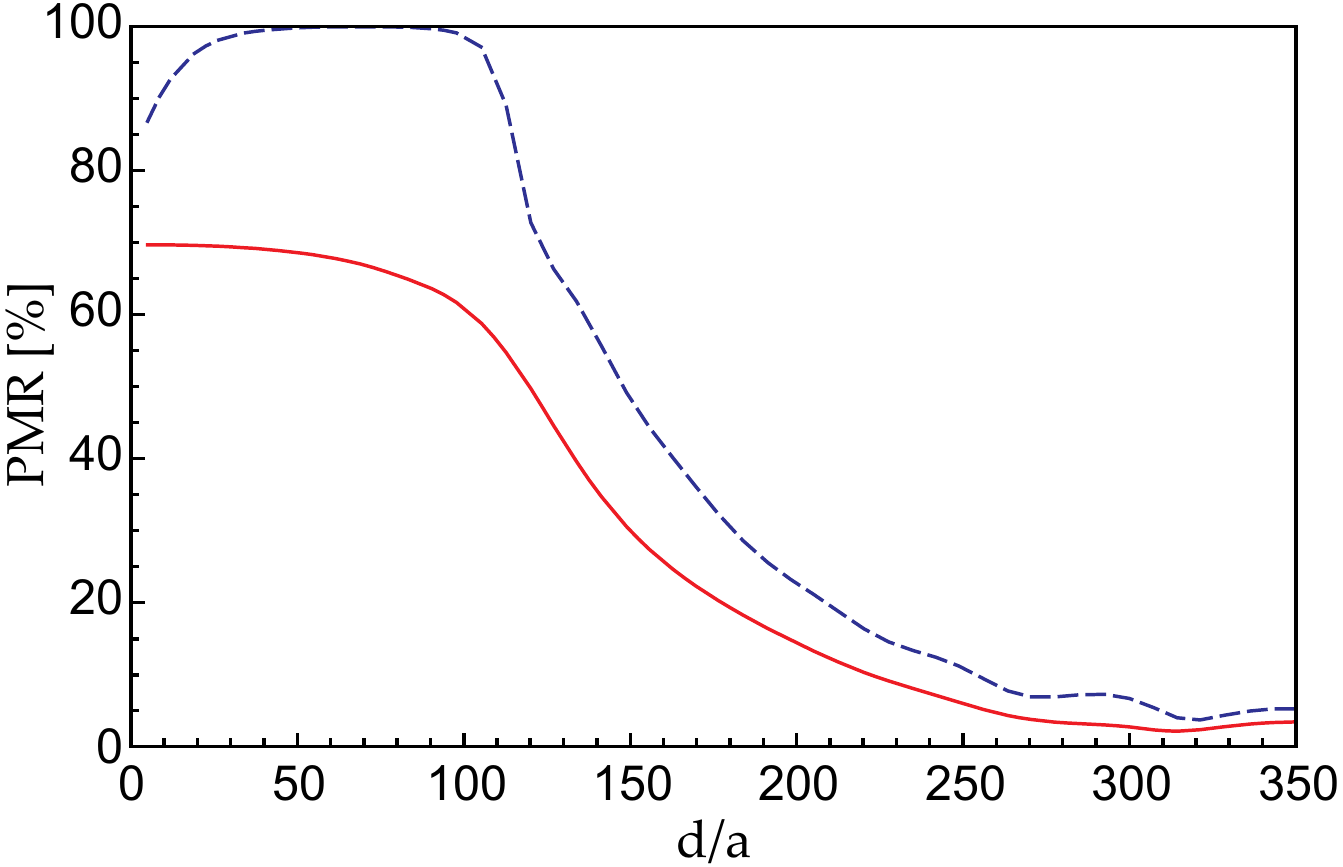}}
\caption{\label{fig:2} Pseudo-magnetoresistance ratio of the
pseudospin valve (solid curve) and the pseudospin-valve transistor
(dashed curve) as a function of the sharpness $d$ of the interface
between regions of different polarity. The Fermi energy is set to
$E_F=U_0=0.07$ eV.}
\end{figure}

In both devices studied, the amount of reflection of incoming
electrons in the anti-parallel configuration will therefore depend
on the sharpness of the interface $d$ as compared to the
pseudospin precession length $l_s$. Figure \ref{fig:2} shows the
computed dependence of the resistance ratio on $d$ for  fixed
values of $E_F=U_0=0.07$ eV. For these parameters, the pseudospin
precession length $l_s\approx 165\,a$. In the regime $d/l_s < 1$
of an abrupt interface, the electron's pseudospin is not able to
rotate quickly enough to accommodate the change, which causes
reflection and a large spin-valve effect. The series connection of
two interfaces in the transistor further increases the resistance
ratio to almost $100\%$, except for very small values of $d$ where
the resistance in the antiparallel arrangement drops due to
tunneling through the central region. In the opposite limit
$d/l_s\gg 1$, the pseudospin of incoming electrons is able to
adiabatically adapt itself to the change of local polarization. In
this limit the pseudospin-valve effect becomes negligible both for
the single interface as well as for the series connection of two
interfaces. The numerical results confirm that the transition
between both regimes occurs at $d\approx l_s$. The
pseudospin-valve effect can therefore be realized in devices with
gate separation $d$ of the order of a few tens of nanometers.

{\em Discussion and Conclusions.---}The large mobility of charge
carriers in graphene has stimulated intense research efforts that
aim at the realization of graphene-based electronic devices. In
particular, manipulation of the differential population of valley
states in momentum space has been proposed \cite{ryc07}, leading
to a ``valleytronic'' analogy of spintronics. Our proposal of
bilayer-based pseudospintronics relies on differential population
of atomic orbitals in real space. This offers a robust mechanism
to exploit spintronic analogies without the necessity of carefully
fabricated nanoribbon edges, which limit the scalability of
valleytronics \cite{falko} and induce harmful intervalley
scattering \cite{akh08}.

In particular, pseudospintronics relies on bulk effects which do
not depend on the crystallographic orientation of the interface.
For the wide samples considered here ($W \gg
\lambda_F\simeq\l_s\gtrsim d$), effects from the sample edges can
be neglected since (i) edge states are localized at realistic
rough edges, (ii) hypothetical clean edges at most contribute an
additional transport channel per spin, and (iii) intervalley
scattering off the edges can contribute to pseudospin relaxation
across the interface, but this effect is negligible for $d\ll W$.
Pseudospintronics is also remarkably robust against bulk disorder.
Chirality guarantees that the bulk pseudospin-flip rate for
majority carriers in a clean ballistic bilayer is zero (the out-of
plane polarization of the pseudospin in the leads is
valley-independent). The predominant scattering mechanism in
graphene, Coulomb scattering off charged impurities, does not
break chirality \cite{moro08,adam08}. Inter-valley scattering
contributes to pseudospin-flip scattering in the interface region,
but the scattering lengths $l_{KK'}\simeq 500{\rm nm}$ reported in
recent experiments \cite{Gorbachev} indicates that this does not
add an additional constraint on $d$.

Additional advantages of the proposed bilayer pseudospintronics
concept arise from the fact that the charge carrier densities are
finite. In devices that involve positioning the Fermi level within
the bandgap, including field-effect transistors \cite{oostinga} or
valley filters based on topologically-confined channels between
insulating regions \cite{martin}, the effective gap size is
reduced by screening \cite{mcc06b,min07}. Under these conditions,
the largest gaps observed so far in experiment are of the order of
$2U_0 \approx 10$~meV~$\approx 100$~K \cite{oostinga}. A finite
charge density admits far larger gaps with experimental values
reaching $2U_0 \approx 200$~meV$\approx 2000$~K
\cite{ohta06,castro} at high density.

We thank V. I. Fal'ko and C. Poole for helpful discussions. This
research was supported by the European Commission, Marie Curie
Excellence Grant MEXT-CT-2005-023778, and by EPSRC First Grant
EP/E063519/1.

\end{document}